\documentclass[12pt,thmsa,a4paper,notitlepage]{article}
\usepackage{amssymb}

\usepackage{sw20aip}

\input{tcilatex}
\begin{document}

\title{Experimental Demonstration of Unconditional Entanglement Swapping for
Continuous Variables}
\author{Xiaojun Jia, Xiaolong Su, Qing Pan, Jiangrui Gao,  \and Changde Xie* and
Kunchi Peng \\
The State Key Laboratory of Quantum Optics and Quantum \\
Optics Devices,\\
Institute of Opto-Electronics, \\
Shanxi University, Taiyuan, 030006, P.R.China}
\maketitle

\begin{abstract}
The unconditional entanglement swapping for continuous variables is
experimentally demonstrated. Two initial entangled states are produced from
two nondegenerate optical parametric amplifiers operating at
deamplification. Through implementing the direct measurement of Bell-state
between two optical beams from each amplifier the remaining two optical
beams, which have never directly interacted with each other, are entangled.
The quantum correlation degrees of 1.23dB and 1.12dB below the shot noise
limit for the amplitude and phase quadratures resulting from the
entanglement swapping are straightly measured.

\end{abstract}

It has been recognized that quantum entanglement is an important resource in
quantum information and computation. Due to utilizing entanglement shared by
sender and receiver together with local operations and classical
communication, various feats of quantum communication, such as quantum
teleportation\cite{one,two,three,four,five} and quantum dense coding \cite
{six,seven}, have been experimentally demonstrated with both discrete and
continuous quantum systems. An other novel and attractive task in quantum
information is entanglement swapping, which means to entangle two quantum
systems that have never directly interacted with each other. The
entanglement swapping of discrete variables has already been achieved
experimentally with single photons\cite{eight}. However, the post-selection
is a standard intrinsic procedure in quantum communication systems of
discrete variables because if no photon is detected, or the two photons
simultaneously entering remote detectors are not in the same basis, the
corresponding time slot has to be ignored and hence does not contribute to
the raw data\cite{nine}. Later, the protocols of unconditional entanglement
swapping for continuous variables (CVs) were theoretically proposed, in
which the determinant squeezed-state entanglement of continuous
electromagnetic field was exploited thus the post-selection is not needed%
\cite{ten,eleven,twelve}. To the best of our knowledge, the entanglement
swapping of CVs has not been experimentally accomplished so far. Thus it
still is a real challenge to realize unconditional entanglement swapping
without post-selection of ''successful'' events by photon detections. In
this letter, we will present the first experimental realization of CV
entanglement swapping.

Fig.1 is the schematic of the experimental setup. The laser is a home made
CW (continuous wave) intracavity frequency-doubled and frequency stabilized
Nd:YAP/KTP ring laser consisting of five mirrors\cite{thirteen}. The second
harmonic wave output at $0.54\mu m$ and the fundamental wave output at $%
1.08\mu m$ from the laser source are used for the pump field and the
injected signal of two nondegenerate optical parametric amplifiers (NOPAs),
respectively. For obtaining a pair of symmetric Einstein-Podolsky-Rosen
(EPR) entangled optical beams, the two NOPAs (NOPA1 and NOPA2) were
constructed in identical configuration, both of which consist of an $\alpha $%
-cut type-$\Pi $ KTP\ crystal and a concave mirror. The front face of KTP\
is coated to be used as the input coupler. The concave mirror (the output
coupler of EPR\ beam) is mounted on a piezoelectric transducer (PZT) for
locking actively the cavity length of NOPA on resonance with the injected
signal at $1.08\mu m.$ Through a parametric down conversion process of type-$%
\Pi $ phase match, an EPR beam with anticorrelated amplitude quadratures and
correlated phase quadratures may be produced from a NOPA operating in the
state of deamplification, that is, the pump field and the injected signal
are out of phase\cite{seven}. The entangled two modes of EPR beam are just
the signal and idler modes produced from the process, which have identical
frequency with the injected signal at $1.08\mu m$ and the orthogonal
polarization with each other\cite{seven}. In the case, the variances of the
sum of amplitude quadratures and the difference of phase quadrature for the
two entangled modes are both smaller than the shot noise limit (SNL) defined
by the vacuum fluctuation. Because the same laser serves as the pump field
and the injected signal source of two NOPAs, the classical coherence between
a pair of EPR beams generating from two NOPAs is ensured.

The two entangled optical modes, $\hat{a}$, $\hat{b}$ and $\hat{c}$, $\hat{d}
$, from NOPA1 and NOPA2 are distributed to Alice and Bob, respectively.
Alice (Bob) divides mode $\hat{a}$ and $\hat{b}$ ($\hat{c}$ and $\hat{d}$)
in orthogonal polarization with polarizing-beam-splitter PBS1 (PBS2).
Initially, Alice and Bob do not share an entangled state. However, we will
see that Alice and Bob can establish the entanglement of mode $\hat{a}$ and $%
\hat{d}$ if they ask Claire for her assistance and send mode $\hat{b}$ and $%
\hat{c}$ to her. Claire performs a joint measurement of mode $\hat{b}$ and $%
\hat{c}$ by the direct detection system of Bell-state, that is, both the
variances of the sum of amplitude quadratures $\left\langle \delta ^2(\hat{X}%
_{\hat{b}}+\hat{X}_{\hat{c}})\right\rangle $ and the difference of phase
quadratures $\left\langle \delta ^2(\hat{Y}_{\hat{b}}-\hat{Y}_{\hat{c}%
})\right\rangle $ are simultaneously measured by using a self-homodyne
detector with two radio frequency(RF) splitters and two (positive and
negative) power combiners to produce the classical photocurrents $\hat{\imath%
}_{+}^c$ and $\hat{\imath}_{-}^c$ \cite{fifteen}. Claire's detection of mode 
$\hat{b}$ and $\hat{c}$ projects mode $\hat{a}$ and $\hat{d}$ on an
inseparable entangled state, the entanglement of which is not changed by any
local operation on mode $\hat{a}$ or $\hat{d}$ as classical displacements%
\cite{eleven,twelve}. However, the entanglement of mode $\hat{a}$ and $\hat{d%
}$ cannot be used or exhibited without information about Claire's
measurement results\cite{eleven}. For exhibiting the entanglement of mode $%
\hat{a}$ and $\hat{d}$, we send the photocurrents $\hat{\imath}_{+}^c$ and $%
\hat{\imath}_{-}^c$ detected by Claire to Bob, where Bob implements the
amplitude-modulation and phase-modulation on a coherent state light $\hat{%
\beta}_0$\ with $\hat{\imath}_{+}^c$ and $\hat{\imath}_{-}^c$ by means of
amplitude (AM) and phase (PM) modulator, respectively. The coherent light $%
\hat{\beta}_0$\ is a part divided from the fundamental wave of the laser
source, thus it has the identical frequency with the EPR beams at $1.08\mu m$%
. The modulated optical mode $\hat{\beta}_0$\ becomes $\hat{\beta}$:

\begin{equation}
\hat{\beta}=\hat{\beta}_0+g_{+}\hat{\imath}_{+}^c+ig_{-}\hat{\imath}_{-}^c
\end{equation}

The parameter $g_{+\text{ }}$and $g_{-}$ describe the amplitude and phase
gain for the transformation from photocurrent to output light field ($g_{+%
\text{ }}=g_{-}=g$ in the experiment for simplification). Then Bob combines
mode $\hat{d}$ and $\hat{\beta}$ at a mirror $M_r$ of reflectivity $R=98\%$.
In this manner the mode $\hat{d}$ is displaced to $\hat{d}^{\prime }$:

\begin{equation}
\hat{d}^{\prime }=\sqrt{R}\left( \xi _2\hat{d}+\sqrt{1-\xi _2^2}\hat{v}_{%
\hat{d}}\right) +\sqrt{1-R}\left[ \hat{\beta}_0+g_{+}\hat{\imath}%
_{+}^c+ig_{-}\hat{\imath}_{-}^c\right] \text{,}
\end{equation}
$\xi _2$ and $\hat{v}_{\hat{d}}$ are the imperfect transmission efficiency
and vacuum noise introduced by losses of mode $\hat{d}$. If in spite of mode 
$\hat{a}$, the process is an quantum teleportation of mode $\hat{b}$ from
Claire to Bob based on exploiting the entanglement between mode $\hat{c}$
and $\hat{d}$\cite{eleven}. Here, mode $\hat{b}$ is the teleported input
state and Claire corresponds to the sender (Alice) in normal teleportation
systems\cite{two}. The initial entangled state, mode $\hat{c}$ and $\hat{d}$%
, serves as the quantum resource used for teleporting the quantum
information of input state $\hat{b}$ from Claire to Bob (the receiver).
Claire's Bell-state detection on mode $\hat{b}$ and $\hat{c}$ collapses
Bob's mode $\hat{d}$ into a state conditioned on the measurement outputs ($%
\hat{\imath}_{+}^c$, $\hat{\imath}_{-}^c$), that is, Claire's joint
measurement on mode $\hat{b}$ and $\hat{c}$ teleports the quantum
information of mode $\hat{b}$ to mode $\hat{d}$ by means of the quantum
entanglement between mode $\hat{c}$ and $\hat{d}$. Hence after receiving
this classical information from Claire, Bob is able to construct the
teleported state $\hat{b}$ via a simple phase-space displacement of the EPR
field $\hat{d}$ \cite{two}. For avoiding the optical loss of mode $\hat{d}$,
which will unavoidably reduce entanglement, in our experiment AM and PM
transform the photocurrents ($\hat{\imath}_{+}^c$, $\hat{\imath}_{-}^c$)
into a complex field amplitude $\hat{\beta}$ firstly, which is then combined
with the EPR beam $\hat{d}$ at the mirror Mr to affect the displacement of $%
\hat{d}$ to $\hat{d}^{\prime }$. To verify that the entanglement swapping
has been accomplished during the process, we measure the quantum
correlations of the sum of amplitude quadratures and the difference of phase
quadratures between mode $\hat{a}$ and $\hat{d}^{\prime }$ with spectrum
analyzers (SAs). If both the quantum fluctuation of the sum and difference
photocurrents are less than the corresponding SNL, the mode $\hat{a}$ and $%
\hat{d}^{\prime }$ are in an entangled state\cite{twelve}. Through analogous
calculation with Refs.[11] and [12], but taking into account the imperfect
detection efficiency of the detectors $(\eta <1)$ and the imperfect
transmission efficiency of the optical system $(\xi _{1-4}<1)$, we can
obtain the noise power spectra of the sum and difference photocurrents. The
calculated variances of the sum and the difference photocurrents are:

\begin{eqnarray}
\left\langle \delta ^2i_{+}^v\right\rangle &=&\left\langle \delta
^2i_{-}^v\right\rangle =\frac 14\left( \eta \xi _3-g_{swap}\eta \xi
_4\right) ^2e^{2r_1}+\frac 14\left( \sqrt{R}\eta \xi _2\xi _4-g_{swap}\eta
\xi _4\right) ^2e^{2r_2} \\
&&+\frac 14\left( \eta \xi _3+g_{swap}\eta \xi _4\right) ^2e^{-2r_1}+\frac
14\left( \sqrt{R}\eta \xi _2\xi _4+g_{swap}\eta \xi _4\right)
^2e^{-2r_2}+1-\eta ^2  \nonumber \\
&&+\frac 12\eta ^2\left( 2-\xi _3^2-\xi _4^2\right) +\frac 12\eta ^2\left(
1-R\xi _2^2\right) \xi _4^2+\frac{g_{swap}^2\left( 1-\eta ^2\xi _1^2\right)
\xi _4^2}{\xi _1^2}\text{,}  \nonumber
\end{eqnarray}
$\xi _1$, $\xi _2$, $\xi _3$ and $\xi _4$ are the transmission efficiency
for mode $\hat{b}$ $(\hat{c})$, $\hat{d}$, $\hat{a}$ and $\hat{d}^{\prime }$%
. $\eta $ is the detection efficiency of each detector, here we have assumed
that the detection efficiency of all detectors (D1-D4) is equal. $%
g_{swap}=\frac 1{\sqrt{2}}\sqrt{1-R}\eta \xi _1g$ is the normalized gain
factor. $r_1$ and $r_2$ are the correlation parameter (also named as
squeezing parameter) for two initial EPR beams from NOPA1 and NOPA2,
respectively. The correlation parameter $r$ is a system parameter depending
on the strength and the time of parametric interaction (nonlinear
coefficient of crystal, pump intensity, finesse of cavity and so on). Under
given experimental conditions, the correlation parameter deciding
entanglement degree is a constant value, thus we say, squeezed-state
entanglement of CV is determinant.

Minimizing Eq.(3) we get the optimum gain factor for the maximum
entanglement:

\begin{equation}
g_{swap}^{opt}=\frac{\eta ^2\left( \left( e^{4r_1}-1\right) e^{2r_2}\xi
_3+e^{2r_1}\left( e^{4r_2}-1\right) \sqrt{R}\xi _2\xi _4\right) \xi _1^2}{
\left[ 4e^{2(r_1+r_2)}+\eta ^2\left(
e^{2r_1}+e^{2r_2}+e^{4r_1+2r_2}+e^{2r_1+4r_2}-4e^{2(r_1+r_2)}\right) \xi
_1^2\right] \xi _4}\text{.}
\end{equation}

Fig.2 is the calculated noise power of $\left\langle \delta
^2i_{+}^v\right\rangle =\left\langle \delta ^2i_{-}^v\right\rangle $
normalized to SNL as a function of the correlation parameters $r_1$ and $r_2$%
, in the numerical calculation $\xi _1^2=0.970$, $\xi _2^2=0.950$, $\xi
_3^2=0.966$, $\xi _4^2=0.968$, $\eta ^2=0.90$, and $R=0.98$ are taken, which
are the real parameters of our experimental system. The dark star designated
in Fig.2 corresponds to the correlation variance deserved with the $r_1$ and 
$r_2$ obtained in the experiment, which is $71.9\%$ of the SNL
(corresponding to $1.43dB$ below the SNL).

In the experiments, at first, we locked both NOPA1 and NOPA2 to resonate
with the injected signal of $1.08\mu m$ from the Nd:YAP/KTP laser and locked
the relative phase between the pump light of $0.54\mu m$ and the injected
signal to $(2n+1)\pi $ (n is integers) for enforcing two NOPAs operating at
deamplification. In this case, the EPR beam of about $70\mu W$ was obtained
from each NOPA at the pump power about $150mW$ just below the power of its
oscillation threshold of about $175mW$ and the injected signal power of $%
10mW $ before entering the input coupler of the NOPA cavity. With respect to
the squeezed vacuum state with average intensity close to zero produced from
an OPA without injected signal\cite{two}, we say, the obtained EPR beam is
bright. The measured correlation degrees of amplitude sum and phase
difference at the sideband mode of $2MHz$ are $\left\langle \delta ^2(\hat{X}%
_{\hat{a}}+\hat{X}_{\hat{b}})\right\rangle =\left\langle \delta ^2(\hat{Y}_{%
\hat{a}}-\hat{Y}_{\hat{b}})\right\rangle $ $=4.10\pm 0.20dB$ below the SNL
for NOPA1 and $\left\langle \delta ^2(\hat{X}_{\hat{c}}+\hat{X}_{\hat{d}%
})\right\rangle =\left\langle \delta ^2(\hat{Y}_{\hat{c}}-\hat{Y}_{\hat{d}%
})\right\rangle $ $=4.30\pm 0.17dB$ below the SNL for NOPA2. Considering the
influence of ENL (electronic noise level), which is 11.3dB below the SNL,
the actual correlations of quadrature components of EPR light beams should
be $4.9dB$ for NOPA1 and $5.1dB$ for NOPA2, respectively.

Substituting the actual correlation parameters of the two initial EPR beams $%
\left( \hat{a},\hat{b}\right) $ and $\left( \hat{c},\hat{d}\right) $, $%
r_1=0.564(4.9dB)$ and $r_2=0.587(5.1dB)$, into Eq.(4), we have $%
g_{swap}^{opt}=0.74$. According to the optimum gain value the classical
channels from Claire to Bob are carefully adjusted in a manner described in
Ref.[5] to the optimum value of $g_{swap}^{opt}=0.74\pm 0.02$. Claire
performs a combining Bell-state measurement of mode $\hat{b}$ and $\hat{c}$
and sends the measured photocurrents $\hat{\imath}_{+}^c$ and $\hat{\imath}%
_{-}^c$ to Bob to modulate a coherent state light $\hat{\beta}_0$. Locking
the relative phase of mode $\hat{d}$ and $\hat{\beta}$ on $M_r$ to $2n\pi $,
the displacement of mode $\hat{d}$ to $\hat{d}^{\prime }$ in the reflective
field is completed. The intensity of $\hat{\beta}_0$ is aligned to make the
intensity of mode $\hat{d}^{\prime }$ equal to that of mode $\hat{a}$ for
satisfying the requirement of Bell-state detection at Victor\cite{fifteen}.
For locking the relative phase between mode $\hat{d}$ and $\hat{\beta}$ to $%
2n\pi $, a PZT is placed in the optical path of $\hat{\beta}$ (not shown in
Fig.1) and the locking technique of DC interference fringe is utilized\cite
{five}.

Successively, Victor implements a direct Bell-state measurement on mode $%
\hat{a}$ and $\hat{d}^{\prime }$. The trace 4 in Fig.3(A) and (B) are the
measured correlation noise powers of the amplitude sum(A), $\left\langle
\delta ^2(\hat{X}_{\hat{a}}+\hat{X}_{\hat{d}^{\prime }})\right\rangle $, and
the phase difference(B), $\left\langle \delta ^2(\hat{Y}_{\hat{a}}-\hat{Y}_{%
\hat{d}^{\prime }})\right\rangle $, at the sideband mode of $2MHz$
respectively, both of which are below the corresponding SNL (trace 3). The
anticorrelation of the amplitude quadratures and the correlation of the
phase quadratures are $1.23dB$ and $1.12dB$ below the SNL, respectively
(after considering the influence of the electronics noise they should be $%
1.34dB$ and $1.22dB$, respectively), which is reasonable agreement with the
calculated value ($1.43dB$). The initial entanglement of mode $\hat{a}$ and $%
\hat{b}$ is $4.9dB$ ($r_1=0.564$), so the percentage of entanglement
preserved after swapping with respect to the initial entanglement value is
about $29\%$. The trace 1 in Fig.3 (A) and (B) are the noise power spectra
of the amplitude sum and phase difference of mode $\hat{a}$ and $\hat{d}%
^{\prime }$ when the classical channels of $\hat{\imath}_{+}^c$ and $\hat{%
\imath}_{-}^c$ from Claire to Bob are blocked, which are much higher than
traces 4 and also the SNL(trace 3). It verifies obviously the conclusion of
Ref.[11] that the entanglement of mode $\hat{a}$ and $\hat{d}^{\prime }$ can
not be exhibited and used without the assistance of Claire's measurement
results. Even the amplitude noise of single mode $\hat{a}$ (trace 2 in (A))
or mode $\hat{d}^{\prime }$ (trace 2 in (B)) is also higher than the
correlation noise of two modes and the SNL. The results are agreeable with
the characteristic of EPR entangled state light\cite{sixteen}. The measured
results show that the entanglement between mode $\hat{a}$ and $\hat{d}%
^{\prime }$ $(\hat{d}),$ which have never interacted with each other, is
truly established.

We achieved the unconditional entanglement swapping due to exploiting the
determinant squeezed-state entanglement initially produced from two NOPAs
pumped by the same laser. For a system with perfect detection and
transmission efficiencies the calculated degree of entanglement on the
swapped pair of modes according to Eq.(3) should be $2.37dB$ which is an
upper boundary set by imperfectly initial entanglement degrees of EPR beams
used for swapping ($4.9dB$ and $5.1dB$). The long-term intensity and
frequency stability of laser source as well as good mechanic and thermal
stabilities of NOPAs are the important requirements for demonstrating the
experiments. This experiment realized the unconditional teleportation of CV
entanglement. Therefore this presented experimental protocol may have
remarkable application potential in quantum communication and computation.

Acknowledgements: We thank J. Zhang and T. C. Zhang for the helpful
discussions on the experimental design and technology. This work was
supported by the Major State Basic Research Project of
China(No.2001CB309304) and the National Natural Science Foundation of
China(No.60238010, 60378014, 10274045).

* Email: changde@sxu.edu.cn

Captions of figures:

Fig.1 Schematic of the experimental setup. Nd:YAP/KTP-laser source;
NOPA-nondegenerate optical parametric amplification; PBS-polarizing optical
beamsplitter; BS-50\% optical beamsplitter; RF-radio frequency splitter; $%
\boxplus $-positive power combiner; $\boxminus $-negative power combiner;
AM-amplitude modulator; PM-phase modulator; D$_{1-4}$-photodiode detector
(ETX500 InGaAs); SA-spectrum analyzer; Mr-98:2 optical beamsplitter

Fig.2 The fluctuation variances of $\left\langle \delta
^2i_{+}^v\right\rangle =\left\langle \delta ^2i_{-}^v\right\rangle $
normalized to the SNL as a function of correlation parameters ($r_1$ and $%
r_2 $) of the initial EPR beams. The dark star corresponds to the
experimental values $r_1=0.564(4.9dB)$, $r_2=0.587(5.1dB)$, where $%
\left\langle \delta ^2i_{+}^v\right\rangle =\left\langle \delta
^2i_{-}^v\right\rangle =0.719(SNL=1)$.

Fig.3 The correlation noise powers resulting from entanglement swapping at
2MHz as a function of time. (A) 1, The noise power of the amplitude sum
without the classical information from Claire; 2, The noise power of
amplitude of mode $\hat{a}$; 3, SNL; 4, The correlation noise power of the
amplitude sum with the classical information from Claire. (B) 1, The noise
power of the phase difference without the classical information from Claire;
2, The noise power of amplitude of mode $\hat{d}^{\prime }$; 3, SNL; 4, The
correlation noise power of the phase difference with the classical
information from Claire. The measurement parameters of SA: RBW(Resolution
Band Width)-10kHz; VBW(Video Band Width)-30Hz.

\end{document}